\documentstyle [12pt,epsf]{article}
\begin{document}
\def\o{\over}
\def\Ar{\rightarrow}
\def\bar{\overline}
\def\r{\gamma}
\def\d{\delta}
\def\a{\alpha}
\def\b{\beta}
\def\n{\nu}
\def\m{\mu}
\def\k{\kappa}
\def\e{\epsilon}
\def\p{\pi}
\def\th{\theta}
\def\om{\omega}
\def\vp{{\varphi}}
\def\Re{{\rm Re}}
\def\Im{{\rm Im}}
\def\ra{\rightarrow}
\def\t{\tilde}
\def\bar{\overline}
\def\l{\lambda}
\def\G{{\rm GeV}}
\def\M{{\rm MeV}}
\def\eV{{\rm eV}}
\baselineskip=24.5pt
\setcounter{page}{1}
\thispagestyle{empty}
 \topskip 0.5  cm
\begin{flushright}
\begin{tabular}{c c}
& March 1999
\end{tabular}
\end{flushright}
\topskip 2 cm
\centerline{\large\bf Prediction of the CP violation in the Phenomenological} 
\centerline{\large\bf Quark-Lepton Mass Matrix Approach}
\vskip 1 cm
\centerline{M. Fukugita$^{1,2}$, M. Tanimoto$^3$ and T. Yanagida$^4$}
\vskip 1 cm
\centerline{$^2$ Institute for Advanced Study, Princeton, NJ 08540, U. S. A.}
\centerline{$^1$ Institute for Cosmic Ray Research, University of Tokyo,
Tanashi, Tokyo 188, Japan}
\centerline{$^3$ Faculty of Education, Ehime University, Matsuyama
790-8577, Japan}
\centerline{$^4$ Department of Physics and RESCEU, University of Tokyo,
Tokyo 113-0033, Japan}
\vskip 1 cm
 
\vskip 2 cm
\noindent
{\large\bf Abstract}

 We explore the phenomenological quark-lepton mass matrices,
  which are devised following the S$_3$ flavour symmetry 
  principle, yet fully consistent with SU(5) gauge models with 
  the Higgs particles of {\bf 5}, {\bf 45} and their conjugates.
  The model contains 10  free parameters altogether.
  When 6 parameters are fixed by  charge 2/3 quark and charged 
  lepton masses, charge($-$1/3) quark masses and all quark 
mixing matrix elements are 
  predicted to be close to experiment, leaving some narrow ranges 
  still as freedom.
  Further specification of 
  3 more parameters (using, $m_d$, Cabibbo angle and  $|V_{cb}|$) suffices
  to fix the quark mixing matrix nearly completely, and all elements 
  come out to be in 
 accurate agreement with experiment. 
  We obtain the CP violation phase as a prediction.

\newpage
\topskip 0.0 cm
 
Grand unification of gauge theories has not given much hint to our
understanding of
the quark-lepton mass spectrum. The most successful among many attempts
is perhaps the ``prediction'' of the mass relations between the charge 
$-$1/3 quarks (referred simply to as down quarks) and the charged leptons
by an introduction of {\bf 45}-plet Higgs in addition to the standard 
{\bf 5}-plet within the SU(5) grand unification \cite{GJ}.
The relation reads,
\begin{equation}
 m_d=3m_e,~~~~~ m_s=\frac{1}{3}m_\mu,~~~~~ m_b=m_\tau ,
\end{equation}
 \noindent 
which is often called Georgi-Jarlskog mass relation.
No successful prediction, however, has been known for 
the charge 2/3 quark spectrum and hence for quark mixing.
Even more difficult is to understand the CP violation phase and
its origin.

The only approach which turns out to be ``successful'' in giving
correct mass-mixing relations for quarks is
an empirical approach, where some discrete symmetry is imposed on the form 
of mass matrices and fix parameters using some quark masses as input
 \cite{Fritzsch1,Models,Hall}. One of the most unsatisfactory
aspect of such approaches was that its consistency is not clear with the
unified gauge model, which anyway we must impose at some level.
If one would impose the compatibility with a gauge model in a 
straightforward way, we are usually led to unwanted relations for 
quark and lepton masses as a remnant of prototype unified gauge models.

We have devised in a previous paper \cite{FTY1} 
phenomenological quark-lepton mass matrices based on
the S$_3$ permutation symmetry principle in a manner fully 
compatible with SU(5) grand unification. This model results in an
approximate Georgi-Jarlskog relation and a mixing angle 
pattern for the quarks, which are in decent agreement with experiment. 
The model
also successfully 
applies to the neutrino mass-mixing problem with bimaximal mixing
as a natural outcome \cite{FTY2}. 
For sake of simplicity of the argument, we have assumed in \cite{FTY1}
the matrix elements to be all real, ignoring  all phases
which could in principle appear therein; also, accurate 
agreement with experiment was not sought for the mixing angles.

In this paper we extend our analysis allowing for full degrees of freedom
of the mass matrices, trying to cure the defects of the previous model.
We have now more parameters, but they are still tightly constrained
within the model, and this leads to a prediction of the CP violating
phase. Here we should quote an earlier work of Fritzsch \cite{Fritzsch2},
who has also derived the CP violating phase in his matrix model approach.


We begin with 
the Yukawa coupling in the SU(5) model:  
\begin{eqnarray}
&&{\cal L}_{\rm Yukawa}= Y(5_H)_{Uij} {\bf 10}_i {\bf 10}_j  {\bf 5}_H +
   Y(45_H)_{Uij} {\bf 10}_i {\bf 10}_j  {\bf 45}_H  \nonumber \\
 && +Y(5_H^*)_{{D/E}ij} {\bf 5}^*_i {\bf 10}_j {\bf 5}_H^* +
   Y(45_H^*)_{{D/E}ij} {\bf 5}^*_i {\bf 10}_j {\bf 45}_H^* + 
     \kappa(5_H 5_H)_{\nu ij} {\bf 5}^*_i {\bf 5}^*_j 
   {{\bf 5}_H{\bf 5}_H \over M_R},
 \label{Lag}
 \end{eqnarray} 
 \noindent                                    
where bald face symbols 
with suffix $H$ denote Higgs scalars of a specified multiplet,
and those with suffix $i$ or $j$ (refer to flavour) 
are SU(5) matter fields,
${\bf 5}_i^* =(d^c_1, d^c_2,d^c_3, e^-, \nu_e)_{Li}$ and
${\bf 10}_j =(u^c_1, ..., u^c_1, ..., d^c_i, ..., e^+)_{Lj}$.
We specify down quarks and charged leptons with suffix $D/E$ as they 
are unified.
 The last term of eq.(\ref{Lag}) is an effective neutrino coupling where 
the neutrino is assumed to be of the Majorana type. 
We suppose that it is induced from heavy Majorana right-handed neutrinos
$N_i$ (SU(5) singlet) with mass $M_R$ \cite{FTY1}, so that ${\bf 45}_H{\bf 45}_H$ 
does not appear in (\ref{Lag}).

We postulate the mass matrices of the form \cite{FTY1}:

 \begin{equation}
 M_D= {K_D \over 3} \Bigg (
\left[ \matrix{1 & 1 & 1 \cr
                           1 & 1 & 1 \cr
                            1 & 1 & 1 \cr
                                         } \right]
+\left[ \matrix{-\e_D & 0 & \e_D \cr
                  0 & \e_D & 0 \cr
                         \e_D & 0 & \delta_D \cr
                                              } \right] \Bigg)  ,
\end{equation}										  
 for the down-quarks, and 
 \begin{equation}
M_E= {K_D \over 3} \Bigg (
\left[ \matrix{1 & 1 & 1 \cr
                            1 & 1 & 1 \cr
                            1 & 1 & 1 \cr
                                         } \right]
+\left[ \matrix{-\e_E & 0 & \e_E \cr
                  0 & \e_E & 0 \cr
                         \e_E & 0 & \delta_E \cr
                                              } \right] \Bigg) ,
\end{equation}		
\noindent
for the charged leptons.
Here, the main part of the mass matrices is induced by a {\bf 5} plet,
which is 
S$_3$ permutation symmetry invariant. We write it as 
S$_3^{\bf 10}\times$S$_3^{\bf 5}$ 
where {\bf 10} and {\bf 5} refer to representations of fermions.
We break S$_3$ symmetry in a hierarchical manner. We introduce
$\delta$ terms to break  S$_3^{\bf 10}\times$S$_3^{\bf 5}$ down to
S$_2^{\bf 10}\times$S$_2^{\bf 5}$.
We assume that $\delta_D$ and $\delta_E$ elements are generated 
from the coupling to a
${\bf 45^*}_H$-plet Higgs scalar. This assignment removes the
unwanted down-quark charged lepton 
mass degeneracy of the minimal SU(5) model, but produces the
Georgi-Jarlskog mass relation. Further symmetry breaking is caused
by $\epsilon$ terms ($\epsilon\ll\delta$)
in a way consistent with ${\bf 5^*}_H$ to allow further adjustment
of mass hierarchy.
We have then
\begin{equation}	
\epsilon_E=\epsilon_D , \qquad   
\delta_E=-3\delta_D .     
\end{equation}	
\noindent
For the up quark masses, respecting the same principle of the symmetry breaking
pattern as with the down quark/charged lepton sector), we write
\begin{equation}
M_U= {K_U \over 3} \Bigg (
\left[ \matrix{1 & 1 & 1 \cr
                            1 & 1 & 1 \cr
                            1 & 1 & 1 \cr
                                         } \right]
+\left[ \matrix{-\e_U & 0 & \delta_U \cr
                         0 & \e_U & \delta_U \cr
          -\delta_U & -\delta_U & 0 \cr  } \right]  \Bigg) . 
 \label{MUmat}
\end{equation}
where the main term and $\epsilon$ terms arise from ${\bf 5}_H$
and $\delta$ terms from   ${\bf 45}_H$.
Note that the texture of
$\delta_U$ 
 in eq.(\ref{MUmat})
 is an unique invariant of S$_2^{\bf 10}\times$S$_2^{\bf 5}$
   among the anti-symmetric ({\bf 45}) mass matrix.

In general, parameters $\e_D$, $\d_D$,	$\e_U$ and  $\d_U$ are complex,
and  we can express	them as 		   
$\e_i=|\e_i| e^{i\a_i}$ and $\d_i=|\d_i| e^{i\b_i}$$(i=U,\  D)$,
whereas we assumed them to be all real in \cite{FTY1}.
We write the matrices in  the hierarchical base \cite{Hi} by applying
an orthogonal transformation in order
to investigate the structure of the phase, 

\begin{equation}
     F^T M_D F\equiv \bar M_D =  {K_D \over 3}
             \left[ \matrix{0 & -{2\o \sqrt{3}}\e_D  & -{1\o \sqrt{6}}\e_D\cr
             -{2\o \sqrt{3}}\e_D & {2\o 3}(\d_D- \e_D) & -{1\o
3\sqrt{2}}(2\d_D+\e_D) \cr
             -{1\o \sqrt{6}}\e_D  &  -{1\o 3\sqrt{2}}(2\d_D+\e_D) &
3+{\d_D\o 3}+{2\e_D\o 3}\cr
         } \right]  ,
\end{equation}	
\noindent and 
\begin{equation}
     F^T M_U F\equiv \bar M_U={K_U \over 3}
             \left[ \matrix{0 & -{1\o \sqrt{3}}\e_U  & -\sqrt{2\o 3}\e_D\cr
             -{1\o \sqrt{3}}\e_D & 0 & \sqrt{2}\d_U \cr
             -\sqrt{2\o 3}\e_U  &  -\sqrt{2}\d_U & 3 \cr } \right]  ,
\end{equation}		
\noindent where
\begin{equation}
F= 
\left[ \matrix{{1\o \sqrt{2}} &  {1\o \sqrt{6}} & {1\o \sqrt{3}} \cr
               -{1\o \sqrt{2}}&  {1\o \sqrt{6}} & {1\o \sqrt{3}} \cr
                      0       & -{2\o \sqrt{6}} & {1\o \sqrt{3}} \cr
                                         } \right]  .
\end{equation}

Extra phases in  $\bar M_D$ and $\bar M_U$ are removed by a phase
transformation applied to
 the quark fields $q_L$ and $q_R$.  With diagonal phase matrices
$P_D$ and $P_U$,
 we obtain
 \begin{equation}
 \widehat M_D = P_D\bar M_D P_D\simeq {K_D \over 3}
             \left[ \matrix{0 & -{2\o \sqrt{3}}|\e_D|  & -{1\o
\sqrt{6}}|\e_D|e^{i\b_D}\cr
       -{2\o \sqrt{3}}|\e_D| & {2\o 3}|\d_D| e^{-i\b_D}& -{\sqrt{2}\o 3}|\d_D| \cr
       -{1\o \sqrt{6}}|\e_D| e^{i\b_D}  &  -{\sqrt{2}\o 3}|\d_D| & 3\cr
         } \right ] ,
	\label{MDh}
\end{equation}	
\noindent and 
\begin{equation}
 \widehat M_U = P_U\bar M_U P_U= {K_U \over 3}
 \left [ \matrix{0 & -{1\o \sqrt{3}}|\e_U|  & -{2\o \sqrt{6}}|\e_U|e^{i\b_U}\cr
             -{1\o \sqrt{3}}|\e_U|& 0 & \sqrt{2}|\d_U| \cr
             -{2\o \sqrt{6}}|\e_U| e^{i\b_U}  &  -\sqrt{2}|\d_U| & 3\cr
         } \right ]  ,
	\label{MUh}
\end{equation}	
\noindent where
\begin{equation}
P_D= 
\left [ \matrix{ e^{-i(\a_D-\b_D)} &  0 & 0 \cr
                0  & e^{-i\b_D} & 0  \cr
                0  & 0 & 1 \cr
                                         } \right ] ,
\qquad 	
P_U= 
\left [ \matrix{ e^{-i(\a_U-\b_U)} &  0 & 0 \cr
                0  & e^{-i\b_U} & 0  \cr
                0  & 0 & 1 \cr
                                         } \right ] .								 
\end{equation}
\noindent
In eq.(\ref{MDh}), only leading terms are retained in the (2,2), (2,3), (3,2)
and (3,3) elements for simplicity of our expressions, while we keep
all terms when we carry out a numerical analysis. The expression of 
eq.(\ref{MUh}) is exact.
Phase matrices $P_D$ and $P_U$ contribute to the Cabibbo-Kobayashi-Maskawa
(CKM) matrix as:
\begin{equation}
 V_{\rm CKM}=U_U^\dagger P_U^\dagger P_D  U_D ,
 \label{VCKM}
\end{equation}
\noindent
where $U_D$ and $U_U$ are unitary matrices, which diagonalize $\widehat M_D$
and $\widehat M_U$.  For convenience, we define the phase matrix $Q$ by
\begin{equation}
  Q=P_U^\dagger P_D =\left [ \matrix{ e^{i\sigma} &  0 & 0 \cr
                0  & e^{i\tau} & 0  \cr
                0  & 0 & 1 \cr    } \right ]  ,
\end{equation}
\noindent
where  $\sigma=(\a_U-\b_U)-(\a_D-\b_D)$ and $\tau=\b_U-\b_D$.

The unitary matrix $U_D$ is given by \cite{Hall}
\begin{eqnarray}
  U_D =&&\left [ \matrix{ 1 & 0 & 0 \cr
                0  & c_2^D & s_2^D  \cr
                0  & -s_2^D & c_2^D\cr } \right ]  
	\left [ \matrix{ e^{-i{\b_D\o 2}} &  0 & 0 \cr
                0  & e^{i{\b_D\o 2}} & 0  \cr
                0  & 0 & 1 \cr  } \right ] 
	\left [ \matrix{ c_1^D & s_1^D & 0 \cr
                -s_1^D  & c_1^D & 0  \cr
                0  & 0& 1 \cr } \right ]  \nonumber \\	
	&& \times\left [ \matrix{ e^{-i{\b_D\o 2}} &  0 & 0 \cr
                0  & 1 & 0  \cr
                0  & 0 & 1 \cr  } \right ] 	
	\left [ \matrix{ c_3^D & 0& s_3^D \cr
                0 & 1 & 0  \cr
                -s_3^D  & 0 & c_3^D \cr } \right ] , 										 
\end{eqnarray}
\noindent
\noindent
where $s^D_i=\sin \theta^D_i$ and $c^D_i=\cos\theta^D_i$ with
\begin{equation}
  s_1^D\simeq -\sqrt{3}{|\e_D|\o |\d_D|}\simeq -\sqrt{m_d\o m_s} , \ 
  s_2^D\simeq -{\sqrt{2}\o 9}|\d_D| \simeq -{1\o \sqrt{2}}{m_s\o m_b} , \ 
  s_3^D\simeq -{|\e_D| \o 3\sqrt{6}}\simeq -{1\o 2\sqrt{2}}\sqrt{m_d m_s\o
m_b^2}.
\end{equation}
\noindent
The down-quark masses are then,
\begin{eqnarray}
   &&m_b\simeq K_D(1+{1\o 9}|\delta_D|\cos\b_D), \nonumber\\
   &&m_s\simeq {2\o 9}K_D |\delta_D|\left (1-{1\o 18}|\d_D|\cos\b_D\right
),\nonumber \\
   &&m_d\simeq -{2\o 3}K_D {|\e_D|^2\o |\d_D|} \left (1+{1\o
4}|\delta_D|\cos\b_D\right ) .
\label{md}
\end{eqnarray}
The charged-lepton masses are given as
\begin{eqnarray}
   &&m_\tau\simeq K_D(1-{1\o 3}|\delta_D|\cos\b_D), \nonumber\\
   &&m_\mu\simeq -{2\o 3}K_D |\delta_D|\left (1+{1\o 6}|\d_D|\cos\b_D
\right ),\nonumber \\
   &&m_e\simeq {2\o 9} K_D {|\e_D|^2\o |\d_D|} \left (1-{3\o 4}
|\delta_D|\cos\b_D\right ) .
   \label{me}
\end{eqnarray}

The matrix for the up quark sector $U_U$ is obtained as \cite{Hall}
 \begin{equation}
  U_U =\left [ \matrix{ 1 & 0 & 0 \cr
                0  & c_2^U & s_2^U  \cr
                0  & -s_2^U & c_2^U\cr } \right ]  
	\left [ \matrix{ c_1^U & s_1^U & 0 \cr
                -s_1^U  & c_1^U & 0  \cr
                0  & 0& 1 \cr } \right ]  
	\left [ \matrix{ e^{-i\b_U} &  0 & 0 \cr
                0  & 1 & 0  \cr
                0  & 0 & 1 \cr  } \right ] 	
	\left [ \matrix{ c_3^U & 0& s_3^U \cr
                0 & 1 & 0  \cr
                -s_3^U  & 0 & c_3^U\cr } \right ] , 										 
\end{equation}
\noindent
where 
\begin{equation}
  s_1^U\simeq -{\sqrt{3}\o 2}{|\e_U|\o |\d_U|^2}\simeq -\sqrt{m_u\o m_c} , \  
  s_2^U\simeq -{\sqrt{2}\o 3}|\d_U| \simeq -\sqrt{{m_c\o m_t}} , \  
  s_3^U\simeq -{2 \o 3\sqrt{6}}|\e_U|\simeq -\sqrt{2}\sqrt{m_u m_c\o m_t^2}.
  \label{md2}
\end{equation}
\noindent
The up-quark masses are 
\begin{equation}
   m_t\simeq K_U \left (1-{2\o 9}|\delta_U|^2 \right), \qquad
   m_c\simeq {2\o 9}K_U |\delta_U|^2 ,\qquad
   m_u\simeq -{1\o 6}K_U {|\e_U|^2\o |\d_U|^2}  .
  \label{mu}
\end{equation}

Substituting  $U_D$ and $U_U$ into  eq.(\ref{VCKM}), we obtain
 the CKM matrix in terms of the quark masses and the phase parameteres.
 After taking $c_1^i\simeq c_2^i\simeq c_3^i\simeq 1(i=D,U)$ and
 neglecting $s_1^i s_3^i$, $s_2^i s_3^i$ and $s_1^i s_2^i s_3^i(i=D,U)$ terms,
 we obtain the CKM matrix elements approximately as

\begin{eqnarray}
&&V_{ud}\simeq \sqrt{1-{m_d\o m_s}}e^{i(\sigma-\b_D)} ,\nonumber \\
&&V_{us}\simeq \sqrt{{m_d\o m_s}}e^{i(\sigma-\b_D)}- \sqrt{{m_u\o
m_c}}e^{i\tau} , \nonumber\\
&&V_{ub}\simeq  -{1\o 2\sqrt{2}}\sqrt{{m_dm_s\o m_b^2}}e^{i(\sigma-\b_D)} 
       -{1\o \sqrt{2}}{m_s\o m_b}\sqrt{{m_u\o m_c}}e^{i\tau}+\sqrt{{m_u\o m_t}}
	   +\sqrt{{2m_um_c\o m_t^2}}e^{-i\b_U} , \nonumber\\
&&V_{cd}\simeq \sqrt{{m_d\o m_s}}e^{i\tau}-\sqrt{{m_u\o
m_c}}e^{i(\sigma-\b_D)}, \nonumber\\
&&V_{cs}\simeq \sqrt{1-{m_d\o m_s}}e^{i\tau} ,\nonumber \\
&&V_{cb}\simeq -{1\o \sqrt{2}}{m_s\o m_b}e^{i\tau}+\sqrt{{m_c\o m_t}} ,
\nonumber\\
&&V_{td}\simeq {1\o 2\sqrt{2}}\sqrt{{m_d m_s\o m_b^2}}+
{1\o \sqrt{2}}{m_s\o m_b}\sqrt{{m_d\o m_s}}
    -\sqrt{{m_d m_c\o m_s m_t}}e^{i\tau}-
	 \sqrt{{2 m_u m_c\o m_t^2}}e^{i(\sigma-\b_D+\b_U)}  , \nonumber\\
&&V_{ts}\simeq {1\o \sqrt{2}}{m_s\o m_b}-\sqrt{{m_c\o m_t}}e^{i\tau} ,
\nonumber\\
&&V_{tb}\simeq 1 .
\label{element}
\end{eqnarray}
\noindent
For the CP violation parameter, the angle $\gamma$ that enters in the unitarity
triangle \cite{Uni} reads,
\begin{equation}
\gamma\simeq-(\sigma-\beta_D)+\sin^{-1}\bigg(-{1 \over 2\sqrt
2}{\sqrt{m_dm_s}\over m_b}
{\sin(\sigma-\beta_D)\over |V_{13}|}\bigg) - \sin^{-1}\bigg( \sqrt{m_u\over m_c}
{\sin(\sigma-\beta_D)\over |V_{12}|}\bigg) ,
\end{equation}
\noindent in the case of a small $\tau$.

These expressions agree with the result of a numerical calculation 
(without any approximations)
within $10 \%$ error.
Although there are four phase parameters $\sigma$, $\tau$, $\b_D$ and
$\b_U$ in our matrices,
 the CKM matrix is determined in practice by only two of them, 
$\sigma-\b_D$ and $\tau$,
 because the last terms of
$V_{ub}$ and $V_{td}$, which contain phases other than the two, are strongly 
suppressed compared with other terms. 
In \cite{FTY1} we have shown that these matrix elements, where all phases
are completely dropped, yield a resonable description for all CKM
matrix elements. The point of the present paper is that all CKM matrix
elements, including the CP violation phase, are completely determined,
once the above two phases are fixed by the adjustment of two of the
CKM matrix element, say  $V_{us}$ and $V_{cb}$.


While proceeding to a  numerical analysis, we need some care as to the 
input data.
Since the CKM matrix elements in eq.(\ref{element}) and masses
eqs.(\ref{md2},\ref{mu})
are discussed at the $SU(5)$
GUT scale,
predictions should also be compared 
at the GUT scale, rather than at the electroweak scale.
Since a supersymmetric extension is the only way to make the SU(5) GUT viable,
we take quark and lepton masses at the GUT scale obtained in 
the minimal SUSY model (MSSM) with the aid of renormalisation group
equation (RGE)
incorporating two-loop\cite{FusaKoi}. These mass parameters are given in
Table 1.

  Let us first discuss charge $-$1/3 quark and charged lepton masses.
  With the three charged lepton masses as input, 
  the parameters in the charged lepton/down quark
  sector are determined to be  $K_D=1.203 {\rm GeV}$, $|\d_D|=0.080$ and
  $|\e_D|=0.0104$, which in fact satisfies $K_D\gg|\d_D|\gg|\e_D|$.
  The ratio of $d$-quark mass to electron mass is given, using 
eqs.(\ref{md}) and 
(\ref{me}), by
 \begin{equation}
   \left |{m_d\o m_e}\right |\simeq 3 \frac{1+{1\o 4}|\delta_D|\cos\b_D}
   {1-{3\o 4}|\delta_D|\cos\b_D}.
 \end{equation}
 \noindent
This shows the dependence of $|m_d/m_e|$ on further two parameters.
By carrying out an accurate numerical calculation with  $|\d_D|=0.08$,
we see that this ratio
takes a value between 2.3 and 3.7, which
corresponds to $m_d=0.76- 1.20 {\rm MeV}$, when $\beta_D$
is varied form 0 to $+\pi$. This is compared with 
the ``experimental value'',
$m_d=1.3\pm 0.2 {\rm MeV}$ (i.e., the ratio is 3.4-4.6). Requiring an
agreement with experiment leads to  $0.7\leq\cos\b_D$. Here
we take $\cos\b_D= 1$ for further analysis. For this case
we have $m_d=1.20 {\rm MeV}$.
The prediction for other down quark masses is given in Table 1.
The values of $m_b$ and $m_s$ are somewhat larger, 
but taken as acceptable when we consider
the uncertainty in the mass analysis using low-order
perturbation theory. 

Now we are concerend with the CKM matrix element.
We obtain the three parameters of the up quark sector to be
 $K_U=129 {\rm GeV}$, $|\d_U|=0.1026$ and $|\e_U|=0.00071$ 
($K_U\gg|\d_U|\gg|\e_U|$ being satisfied) if
   the central values in Table 1 are adopted for up-quark masses.
We note, however, a large uncertainty in the estiamte of the top quark 
mass at the GUT scale, arising from the fact that the top quark mass is
 near the fixed point of RGE  at the electroweak scale.
The prediction of the CKM matrix depends on top quark mass 
used as input;
   we take account of this large error, $89-325{\rm GeV}$
(for a given pole mass $m_t({\rm pole})=180\pm 12 \G$), 
for further analysis of the matrices.  
As with the case of the mass, we must take account of the running effect
of the CKM matrix elements. It is known that
the elements $V_{ud}$, $V_{us}$, $V_{cd}$, $V_{cs}$ and  $V_{tb}$  are nearly 
 constant during running between the electroweak and the GUT scales.  
 On the other hand, all others are affected by the large Yukawa 
coupling of the
top quark by 10-20\%.
We may use $|V_{us}|=0.217-0.224$ (GUT scale) to constrain the phase
$\phi=\sigma-\tau-\b_D$. We then obtain $\phi=(\pm 60.5^\circ)-(\pm 68.5^\circ)$.
Another phase $\tau$ is fixed by  $|V_{cb}|=0.0347-0.036$;
we obtain $\tau=0- (\pm 22)^\circ$ for $m_t=129{\rm GeV}$ at the GUT scale.
All parameters are now fixed, and a prediction is given
for the full CKM matrix elements, as presented in Table 2. 
In this table we also give experimental values and estimates at the
GUT scale after running \cite{FusaKoi}. All predictions are within
the uncertainty of the empirical values. The agreement with experiment
is not spoiled even if we take the upper limit value 
$m_t=325$ GeV at the GUT scale.  On the other hand, if we decrease the top
quark mass
the agreement is disturbed for $|V_{cb}|$: the prediction goes
out of the upper limit of the range 
allowed by experiment, $0.030-0.036$. This limits
our consideration to $123\G \leq m_t \leq 325 \G$.

The most interesting prediction is perhaps that of the CP violation
parameter. We obtain 
\begin{equation}
            \gamma=106^\circ- 114^\circ,
\end{equation}
\noindent
for $m_t=129$ GeV.
  This is equivalent to the vertex position of 
  $\rho=-0.098\sim -0.086$ and $\eta=0.221\sim 0.309$ in 
Wolfenstein's $\rho-\eta$ plane \cite{Wolfen}.
 \footnote[1]{These predictions should in principle be compared with 
 the corresponding values at the GUT scale.
 Fortunately, the energy scale dependence of these quantities is
 very weak  \cite{FusaKoi}, and we can safely neglect
 the running effect.}
If the range of the input $m_t$ is relaxed as discussed above, the
prediction for $\gamma$ becomes $76^\circ-114^\circ$. 
This value is similar to
the prediction of Fritzsch \cite{Fritzsch2}, $\gamma=72^\circ\sim 76^\circ$.
 In Table 3, we present the prediction of the several key parameters 
 for various top-quark masses. Here, we take $|V_{ub}/V_{cb}|$ as
one of the indicators.
Fig. 1 is a  summary of our prediction for the CKM matrix in the
$\rho-\eta$ plane, where currently available experimental constraints 
are also plotted: (i) $\epsilon_K$ parameter with $B_K=0.6-0.9$
 (ii) $|V_{ub}/V_{cb}|=0.08\pm 0.02$ and (iii) $\Delta M_{B_d}$ with
  $\sqrt{B_{B_d}} F_{B_d}=200\pm 40 \M$.

 We have explored the full content of the
 phenomenological quark-lepton mass matrices,
  which are derived from  S$_3$ flavor symmetry principle 
and its hierarchical breaking in the framework of 
  SU(5) gauge models, proposed in \cite{FTY1}.
  Our mass matrices have 10 free parameters altogether,   
  6 real $K_i$, $|\d_i|$, $|\e_i|(i=D,U)$ and
  4 phases $\b_D$, $\sigma$, $\b_D$, $\tau$, excluding neutrino parameters.
Using ($m_u$, $m_c$, $m_t$) and  ($m_e$, $m_\mu$, $m_\tau$) as input, 
we have a prediction in decent agreement with experiment for
 the rest of the physical parameters, irrespective of the free
parameters left unspecified. Requiring more precise agreement for
$m_d$ and two mixing angles, say $|V_{us}|$ and $|V_{cb}|$,
we have a complete determination of the CKM matrix elements including
phases. The obtained matrix shows an excellent
agreement with experiment
within the present experimental accuracy.  
The most interesting among others is 
the prediction of the CP violation phase, which would 
 soon be tested to higher accuracy with B-factories. 

We have also studied the mixing problem for the lepton sector
including neutrinos in a way
parallel to the quark sector, but have found little to add to the 
previous work of ref.\cite{FTY2}, except for phases. With the allowed 
parameter range the net CP
violating phase, as defined by Jarlskog \cite{Jal}, is as small as 
$J_{\rm CP}<10^{-4}$ because the CP violating phase enters only the
 symmetry breaking terms.
This is too small to arouse any phenomenlogical interests.

\vskip 20 mm
\noindent
{\bf Acknowledgements}

MF thanks the Raymond and Beverly Sackler Fellowship and the
Alfred P. Sloan Foundation for the support for the work in
Princeton. MT and TY are supported in part by the Grants-in-Aid of the 
 Ministry of Education of Japan (Nos.10640274, 7107).
\newpage

\newpage
\begin{table}
\hskip -0.5  cm
\begin{tabular}{ | r| r| r| r| r| r| r| r| r| r| } \hline
        &    & &  &      &        &         &        &        &               \\
 & $m_e\ \ $ & $m_\mu\ \ $ &  $m_\tau \ \ $  &  $m_d\quad $ & $m_s\quad$ &
  $m_b \qquad$ &  $m_u \quad$   & $m_c\quad$ &  $m_t\quad$  \\
 & (MeV)       & (MeV)   & (GeV)    &  (MeV)\ \      &  (MeV)\ \   &  (GeV) \ \ \ \  
 &  (MeV)\ \    & (GeV)\     & (GeV)\  \\
 & & & & & & & & &  
\\ \hline
        & & &       &        &        &          &        &     &    \\
 ``exp.'' & $0.325$ & $68.60$  & $1.171$ & $1.3\pm 0.2$ & 
$26.5^{+3.4}_{-3.7}$  
 & $1.00\pm 0.04$ & $1.0\pm 0.2$ & $302^{+25}_{-27}$ & $129^{+196}_{-40}$ \\
  &      &   &   &    &      &       &     &    &   \\  \hline
  & & &     &      &        &          &        &      &        \\ 
pred. & input  & input & input & 1.20\ \ \  \  & 19.6\ \ \ & 1.22\ \ \ \ \ &
input\ \ \
&input\ \ & input\ \ \\
     &      &   &   & (input)  &         &          &    &   &  \\  \hline
\end{tabular}
\caption{Input quark-lepton mass parameters and the prediction of our
model at the GUT scale.
$m_d$ is the value with $\cos\b_D=1$.}
\end{table}
\begin{table}
\hskip 0.1 cm
\begin{tabular}{ | r| r| r| r| } \hline
                &               &        &          \\
 & ``exp.''\ \ \ \ \ \ \ \    &   value at GUT scale   &    prediction at GUT
scale   \\ \hline
	        &           &                        &          \\
  $|V_{ud}|$  &  $0.975-0.976$\ \ & $0.975 - 0.976$\ \  & $0.975 -
0.976\qquad $ \\
  $|V_{us}|$  &  $0.217-0.224$\ \  & $0.217 - 0.224$\ \ & $\underline{0.219
- 0.221}\qquad$ \\
  $|V_{ub}|$  &  $0.0018-0.0045$\ \ & $0.0015 - 0.0040$\ & $0.0019 -
0.0026\qquad$ \\
  $|V_{cd}|$  &  $0.217-0.224$  \ & $0.217 - 0.224$\ \ & $0.219 -
0.220\qquad$ \\
  $|V_{cs}|$  &  $0.974-0.976$\ \ & $0.974 - 0.976$\ \ & $0.975\qquad\qquad$  \\
  $|V_{cb}|$  &  $0.036-0.042$\ \ & $0.030 - 0.036$\ \ & $\underline{0.035
- 0.036}\qquad$ \\
  $|V_{td}|$  &  $0.004-0.013$\ \ & $0.0035 - 0.011$\ \ & $0.007 -
0.008\qquad$\\
  $|V_{ts}|$  &  $0.035-0.042$\ \  & $0.030 - 0.036$\ \ & $0.034 -
0.035\qquad$ \\
  $|V_{tb}|$  &  $0.999\qquad$ & $0.999 - 1.000$\ \ & $0.999\qquad\qquad$\\
  &         &         &               \\  \hline
\end{tabular}
\caption{The CKM matrix elements. The first column shows experiment, the second
is the values estimated at the GUT scale. The third column is the prediction 
of our model with $m_t=129 \G$ at  GUT scale. 
  The underlined values are input.}
\end{table}
\vskip 5 cm

\vskip 5 cm
\newpage
\begin{table}
\hskip 0.1 cm
\begin{tabular}{ | r| r| r| r| r| r| r| } \hline
                &               &        &                 &        &      
 &               \\
$m_t({\rm GeV})$&$|\tau|(^\circ)$ & $|V_{cb}|\quad\ \ $ &
$|V_{ub}/V_{cb}|$\ \ & 
                   $\rho\qquad\quad$\ \  & $\eta$ \qquad \quad&$\gamma
(^\circ)$ \ \\
                &               &        &                 &        &      
 &          \\ \hline
                &               &        &                 &        &      
 &           \\
 129\ \ \ \ & 0-22  & 0.035-0.036 & 0.055-0.072 & $-$0.098 - $-$0.086 &
0.221-0.309 & 106-114\\ 
 150\ \ \ \  & 0-43  & 0.031-0.036 & 0.056-0.085 & $-$0.116 - $-$0.045 &
0.216-0.377 & 97-118\\
 200\ \ \ \  & 41-69 & 0.030-0.036 & 0.091-0.094 & $-$0.068 - $-$0.033 &
0.402-0.421 & 86-100\\
 250\ \ \ \ & 60-85 & 0.030-0.036 & 0.096-0.099 &  0.004 -  0.082\ \ &
0.425-0.445 & 79-90 \\
 325\ \ \ \  & 76-90 & 0.030-0.033 & 0.099-0.103 &  0.068 -  0.105\ \ &
0.435-0.457 & 76-82\\
     &         &         &          &    &   &  \\  \hline
\end{tabular}
\caption{Predictions of the CKM matrix elements and the unitarity triangle.
$m_t$ is the value at the GUT scale.}
\end{table}
\vskip 5 cm
\begin{figure}
\epsfxsize=12 cm
\centerline{\epsfbox{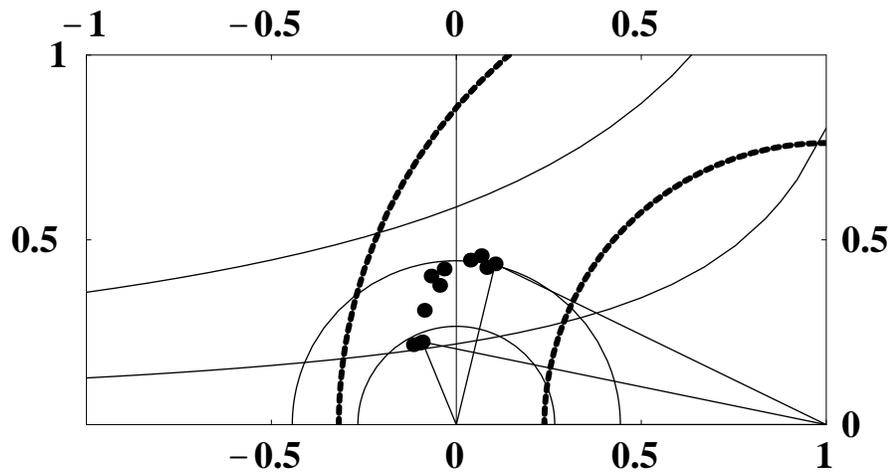}}
\caption{Predicted  vertex of the unitarity triangle
   on the $\rho-\eta$ plane for $m_t({\rm GUT})$ varying between 123$-$325 GeV.
 Experimental constraints from
 $\epsilon_K$, $|V_{ub}/V_{cb}|$ and $\Delta M_{B_d}$
 are also plotted.}
\end{figure}

\end{document}